# Radiation-induced softening of Fe and Fe-based alloys during in-situ electron irradiation under mechanical testing

V.I. Dubinko[1], D. Terentyev[2], A.N. Dovbnya[1], V.A. Kushnir[1], I.V. Hodak[1], S.V. Lebedev[3], S..A. Kotrechko[4], A.V. Dubinko[2,5]

[1] National Science Center "Kharkov Physical-Technical Institute, Kharkov 61108, Ukraine

[2] SCK•CEN, Nuclear Materials Science Institute, Boeretang 200, Mol, 2400, Belgium

[3] V.N. Karazin Kharkov National University, Kharkov 61077, Ukraine

[4] G.V. Kurdyumov Institute for Metal Physics NAS of Ukraine, Kyiv 03142, Ukraine

[5] Ghent University, Applied Physics EA17 FUSION-DC, St.Pietersnieuwstraat, 41 B4 B-9000, Gent, Belgium

**Summary**

Defects formed under irradiation in the bulk act as additional pinning centers resulting in the well-known effect of *radiation-induced hardening*. On the other hand, there is a poorly understood but well-established effect of instant and reversible softening of metals subjected to various types of irradiation. This *radiation-induced softening* (RIS) effect should be taken into account both in the theory of radiation effects and in the engineering approach for technological applications. In the present paper, the RIS is investigated experimentally in polycrystalline technically pure iron (0.048%C) and in commercial ferritic steel. The effect of the electron beam on plastic deformation of bcc Fe is compared with that in fcc Al (99.5%). Electron energy ranged from 0.5 to 0.8 MeV and the beam density ranged from 2.4 to $5 \times 10^{13}$ $cm^{-2}s^{-1}$. Reversible drop of the yield stress and radiation-induced reduction of the elongation to fracture are measured as functions of the electron current and specimen thickness. Rate theory of RIS is proposed, which takes into account the radiation-induced excitation of *moving discrete breathers* (DBs), recently proven to exist in bcc Fe, and their interaction with dislocations enhancing unpinning from structural defects. The behavior of the DBs is studied using classical MD simulations providing input for the modified rate theory, which eventually demonstrates a reasonable agreement with experimental data. The relevance of results to the in-reactor behavior of pressure vessel steels is discussed.

## 1 Introduction

One of the main factors limiting the service lifespan of the nuclear reactors is the radiation embrittlement caused by the radiation-induced hardening of Fe based steels used as structural materials for pressure vessels etc. Defects formed by irradiation in the bulk act as additional pinning centers, resulting in the well-known effect of *radiation-induced hardening*. On the other hand, there is experimental evidence of *radiation-induced softening* (RIS) under low and medium temperature electron, gamma or neutron irradiation. The RIS has been discovered in the early 1960s [1] and investigated extensively thereafter (see e.g. [2]). Single crystals of Zn, Sn, In and Pb have been irradiated at liquid nitrogen temperature (78 K) with electron flux density ranging from $10^{17}$ $m^{-2}s^{-1}$ to $10^{18}$ $m^{-2}s^{-1}$ and energies below and above the threshold displacement energies, the latter being 0.7 MeV (Zn), 0.8 MeV (Sn, In) and 1.2 MeV (Pb). At such low temperatures plastic strain occurs via dislocation glide, the rate of which is limited by thermally activated unpinning of dislocations from local obstacles. The underlying mechanisms of RIS are still a subject of debate [3]. In the first theoretical works on the subject [3], kinetic mechanisms were proposed for the description of RIS under low-temperature irradiation based on the interaction of dislocations with focusons that were suggested to transfer vibrational energy to dislocations facilitating their unpinning from obstacles. Subsequently, new experimental evidence was obtained on the radiation-induced increase of plasticity of polycrystalline Cu (99.5%), Al (99.5%) and Al-3Mg under *in situ* electron irradiation above the room temperature [4]. The electron energy of 0.5 MeV used in these experiments was higher than the threshold displacement energy in Al (0.15 MeV) and about that for Cu. In all cases, irradiation resulted in the decrease of yield stress and increase of the elongation to fracture, i.e. a metal under irradiation instantly becomes *less hard* and *more ductile* as compared to the state prior and after irradiation. These results demonstrated that mechanical properties of materials *under reactor conditions* could be different from those tested 'out of pile' in the surveillance programme.

The present experiments were designed so that to allow comparison between over-threshold irradiation of Al and polycrystalline Fe (and commercial steel 20), which is a base metal of the reactor structural materials in order to make the results more closely related to the real in-reactor environment. For this purpose the beam energy for the Fe irradiation was increased from 0.5 MeV to 0.8 MeV, which is sufficient to produce displacement damage at a rate of $10^{-9}$ dpa/s (where dpa denotes 'displacements per atom') which is comparable to the dose rates in nuclear reactor environment.

The main result obtained here is that the yield stress of Fe is decreased by irradiation, as well as the ultimate resistance to fracture, the latter being in a marked contrast to Al and Cu cases, which points out at the radiation-induced localization of plastic strain in Fe. So the Fe-based metal under irradiation became *less hard* and *more brittle* at the same time. These results mean that dynamics of dislocations has been changed due to their interaction with radiation-induced excitations of the lattice, the nature of which needs to be determined. Focusons proposed in ref. [3] are thermally unstable and they can hardly be responsible for RIS at elevated temperatures. However, in crystals with sufficient anharmonicity, a special kind of lattice vibrations, namely, *discrete breathers* (DBs), also known as intrinsic localized modes, can be generated either by thermal fluctuations or by external triggering. The amplitude of atomic oscillations in the DBs greatly exceeds that of harmonic oscillations (phonons) [5-15]. Due to the crystal anharmonicity, the frequency of atomic oscillations increases or decreases with raising the amplitude so that the DB frequency lies outside the phonon frequency band, which explains the weak coupling of DBs with phonons and, consequently, their stability against decay even at elevated temperatures. DBs have been successfully observed experimentally in various physical systems [8] and materials ranging from metals to diatomic insulators [9], and they have been proposed recently as *catalyzers for various chemical reactions* in solids [16-19]. This field of research is comparatively new, lying at the conjunction of nonlinear physics with material science. The main theoretical result of the present paper is that DBs present a viable *catalyzing mechanism* for the dislocation unpinning from obstacles under irradiation triggering DB generation.

The paper is organized as follows. In the next section, experimental results on electron irradiation of bcc Fe (99.5%) are presented in comparison with analogous results for fcc Al reported in ref. [4]. In section 3, the DB properties and their interaction with edge and screw dislocations in bcc Fe are analyzed based on results of large scale molecular dynamics (MD) simulations using realistic many body interatomic potentials for bcc Fe. In section 4, a rate theory of DB excitation under thermal equilibrium and external driving is presented. In section 5, a model for amplification of the dislocation unpinning from obstacles by radiation-induced DBs is proposed and compared with experimental data. The results are discussed in section 6 and summarized in section 7.

## 2   Experimental investigation of RIS in Fe

### 2.1   *Experimental setup*

The present technical approach combines electron irradiation (compact electron linear accelerator, E< 1 MeV) with in-situ mechanical testing at the installation, which measures a yield stress drops during irradiation pulses and the stress-strain deformation curve under continuous irradiation. Experimental procedure and installation is described in details in ref. [20]. Electron beam of the energy ranging from 0.5 to 0.8 MeV and the beam density ranging from 2.4 to $3.6 \times 10^{13}$ $cm^{-2}s^{-1}$ was directed at a metal specimen subjected to tensile load. The time diagram of the electron beam pulses (the fine pulse structure) is shown in Fig. 1. Micropulses of duration, $\tau_m = 4 \cdot 10^{-11}$s, were shot periodically with an interval of $3 \cdot 10^{-10}$ s during the bunch time, $\tau_{bunch} = (2-4) \times 10^{-6}$s. The bunch frequency, $1/T_0$, was 25 Hz. An overall irradiation time ranged from 10 to 3000 s in different irradiation regimes.

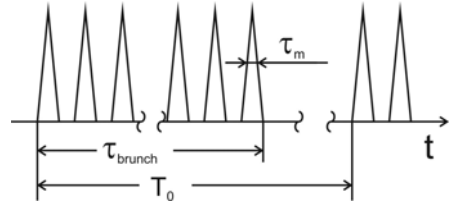

Figure 1. Time dependence of the electron beam pulses [20].

Specimens had a form of a parallelepiped with broader ends for fixing and the following dimensions of the irradiated part: $0.52(\pm0.01) \times 4 \times 30$ mm$^3$. They were cut of technically pure iron (0.048%C) annealed in vacuum in two regimes: (1) T = 900 $^0$C for 1 hour resulting to a mean grain size of 7 microns; (2) T = 1150 $^0$C for 2.5 hours resulting to a mean grain size of 150 microns. At the experimental temperature (slightly above RT), Fe matrix has a bcc structure with impurities precipitated in the form of various obstacles for dislocation glide, which increases the initial yield stress of material (prior to irradiation) to $\sim$ 200 MPa as compared to $\sim$50-100 MPa for technologically pure Fe. The microstructure of the 'steel 20' is much more complex, but principally, it consists of ferritic bcc iron grains (with some Carbon in solution) and a certain fraction of perlite grains, which doubles its strength as compared to the strength of technically pure iron. .

The specimens were subjected to uniaxial tensile load in the deformation installation, which was registered in the coordinates – load, $P$, vs. time, $t$, with delay ranging from 1 to 0.3 s and sensitivity of 0.1%. The load is related to the external stress, $\sigma$, as $\sigma = P(1+\varepsilon)/S$, where $S$ is the specimen cross-section and $\varepsilon$ is the deformation calculated by $\varepsilon = v_d \cdot t/l$, where $v_d$ is the velocity of the deformation rod, $l$ is the specimen length. The load measurement accuracy was about (0.1-1) N. The velocity of the deformation rod was 0.5 $\mu$m$\cdot$s$^{-1}$, which corresponded to the deformation rate of $\dot{\varepsilon} \cong 2 \times 10^{-4}$ s$^{-1}$, i.e. typically used in the standard tensile tests. The surface temperature of the specimens during testing was measured independently with infrared pyrometer and thermocouples attached to the specimen outside the irradiated area.

Effect of the irradiation on plastic deformation of Fe and Al was studied at room temperature applying the discrete and continuous regimes of irradiation. In the first case, specimens were irradiated under external load within the short time intervals $t_{irr}$ followed by the intervals without irradiation. In the second case, the specimens were irradiated under external load continuously up to the fracture point.

### 2.2 Discrete irradiation regime

In the following tests, the specimens were exposed to discrete irradiation pulses with $t_{irr} \sim 60$ s, during which the yield stress drops sharply by the value, $\delta\sigma_\varphi$, and then it increases at a lower rate than that without irradiation, as can be seen in Fig. 2.

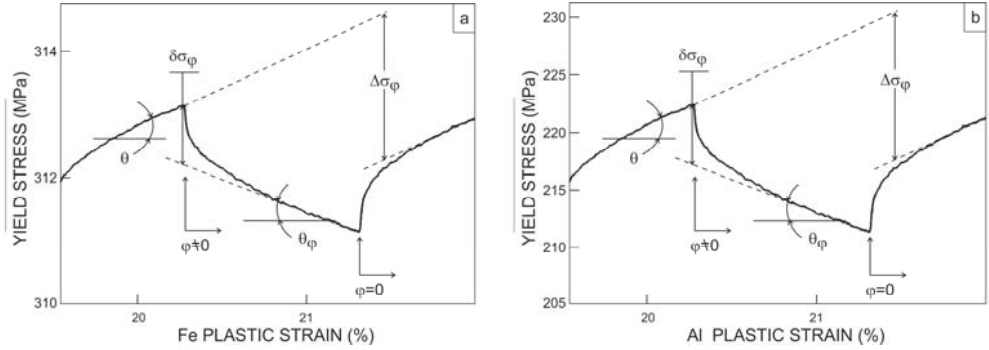

Figure 2. Yield stress drops during deformation (a) of Fe specimens ($T_{ann}$=1150[0] C) under electron beam with E = 0,8 MeV, $\varphi$ = 3.6 $\cdot 10^{13}$ cm$^{-2}$s$^{-1}$ and (b) Al specimens under electron beam with E = 0,5 MeV, $\varphi$ = 5 $\cdot 10^{13}$ cm$^{-2}$.s$^{-1}$ [20].

Let us define a deformation strengthening rate as $\theta = d\sigma/d\varepsilon$. When the electron beam is switched on ($\varphi \neq 0$), the yield stress drops sharply by the value, $\delta\sigma_\varphi$, and subsequently a prolonged deformation stage occurs, at which the deformation strengthening rate $\theta_\varphi$, is always lower than that without irradiation $\theta_0$, and it can be even negative for some time. When the electron beam is switched off ($\varphi = 0$), the yield stress jumps up sharply by the value, $\delta\sigma_\varphi$, (equal to the initial stress drop) and subsequently grows with time at a rate $\theta_0$. As a result, the net external stress decreases by the value $\Delta\sigma_\varphi$, which indicates that the metal microstructure changes during irradiation, the material becomes more soft *under* irradiation pulse as compared to the unirradiated state, and the effect persist for some time after the pulse.

The initial stress drop/jump at the moment of the beam switching on/off, $\delta\sigma_\varphi$, is shown in Fig. 3 for Fe specimens annealed at $T_{ann}$=900° C, which are harder initially than those annealed at 1150° C. It can be seen that $\delta\sigma_\varphi$ increases linearly with increasing deformation and the beam density.

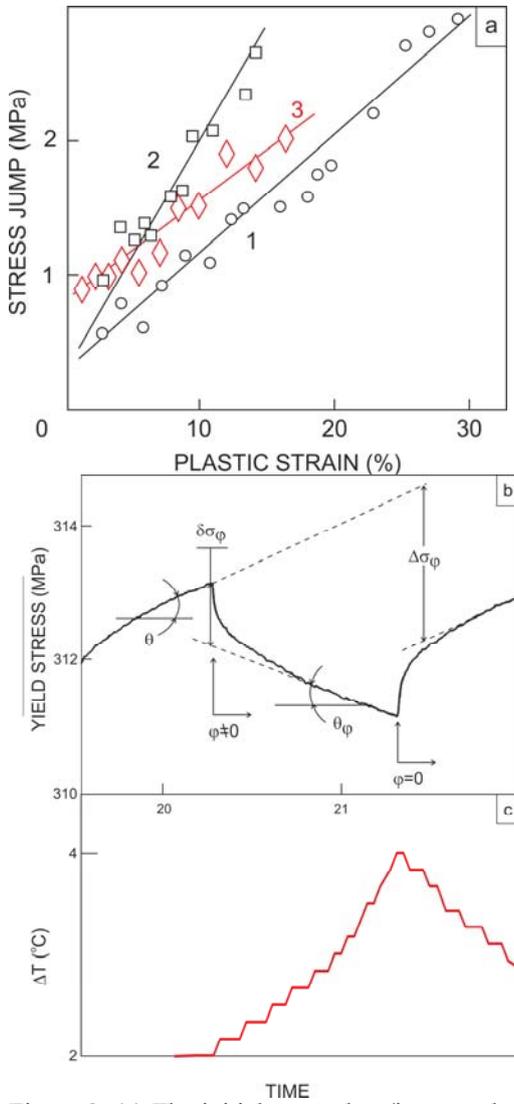

Figure 3. (a) The initial stress drop/jump at the moment of the beam switching on/off as a function of deformation for Fe specimens ($T_{ann}$=900$^{\circ}$C) under electron beam with E = 0,8 MeV, $\varphi$ = 2.4 $\cdot 10^{13}$ cm$^{-2}$s$^{-1}$ (line 1); 3.6 $\cdot 10^{13}$ cm$^{-2}$s$^{-1}$ (line 2); Fe specimens annealed at 1150$^{\circ}$C at $\varphi$ = 3.6 $\cdot 10^{13}$ cm$^{-2}$s$^{-1}$ (line 3, color online). Yield stress evolution of Fe specimens ($T_{ann}$=1150$^{\circ}$C) during one irradiation cycle is shown in (b) while the corresponding time evolution of specimen temperature is shown in (c).

The value $\delta\sigma_{\varphi}$ characterizes an instant (and reversible) response of the metal to irradiation, which can not be explained by the radiation or strain induced transformation of microstructure that takes much longer times than the times of the stress drops/jumps by $\delta\sigma_{\varphi}$. We may conclude here that during irradiation pulses material becomes softer by two mechanisms. One is reversible ($\delta\sigma_{\varphi}$>0) and the other one is irreversible ($\theta_{\varphi}$ <$\theta_{0}$), which results in material softening both during and shortly after the irradiation pulse.

### 2.3 Continuous irradiation regime

During continuous irradiation under increasing strain, the specimen temperature increases gradually as shown in Fig. 4. The deformation curves measured without and under irradiation show that irradiation results in a decrease of the yield stress as well as of the ultimate elongation before fracture of all specimens under investigation. It means that the plasticity limit (strain to fracture) is *decreased* under irradiation by ~20%, in a remarkable contrast to Al and Cu samples for which it was *increased* by ~25-30% [4]. The result demonstrated in Fig.4(a) was reproduced using ??? specimens, every time showing significant reduction of the strain to fracture. Clearly, the capacity of the material to sustain plastic deformation is strongly affected by the irradiation. Let us discuss possible reasons for the reported effects of the irradiation.

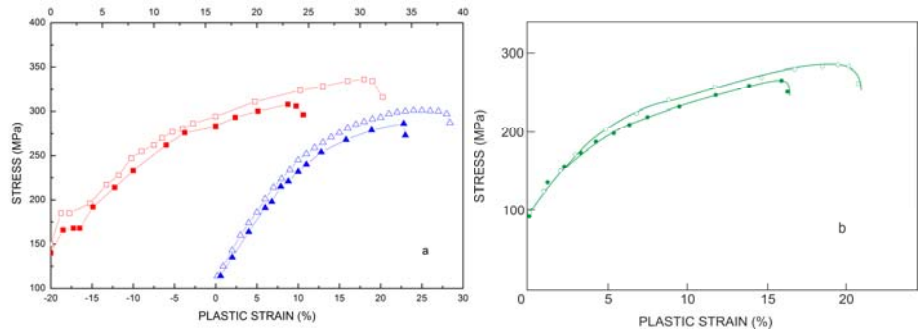

Figure 4. (a) Deformation curves without (open symbols) and under (filled symbols ) irradiation for Fe specimens (red $-$ T$_{ann}$=900$^{\circ}$ C; blue - T$_{ann}$=1150$^{\circ}$ C). Electron energy E = 0.8 MeV, φ = 2.4×10$^{13}$ cm$^{-2}$s$^{-1}$. (b) The same for steel 20 and E = 0.5 MeV, φ = 2.4×10$^{13}$ cm$^{-2}$s$^{-1}$.

### 2.4 RIS dependence on the activation volume

In a phenomenological theory of *thermally activated* dislocation motion [21], the plastic strain rate, $\dot{\varepsilon}_T$, is given by:

$$\dot{\varepsilon}_T = \dot{\varepsilon}_0 \exp\left(-\frac{E_a - V^*\sigma}{k_B T}\right) , \quad \dot{\varepsilon}_0 = bl_d \rho_d \omega_d^0 , \tag{1}$$

where $E_a$ is the activation energy of slip (i.e. energy barrier to activate dislocation motion characterized by specific activation volume, $V^*$, associated with a size of obstacles obstructing dislocation motion), $\sigma$ is the resolved shear stress, $k_B$ is the Boltzmann constant, T is the temperature. The pre-exponential factor can be expressed via material parameters: $b$ - the Burgers vector length, $\rho_d$ - the dislocation density, $l_d$ - the mean length of dislocation segments pinned on the obstacles, and $\omega_d^0$ - the frequency of the dislocation oscillations.

The activation volume can be approximately taken as $V^* \sim b^2 L_d$, where $L_d$ is the mean spacing between the dislocation pinning centers, which decreases with increasing deformation.

Most importantly, it can be measured experimentally both under and without irradiation, as shown in Fig. 5. One can see that the activation volume in Al is higher than that in Fe by a factor of ~3, and both are inversely proportional to the applied deformation, ε.

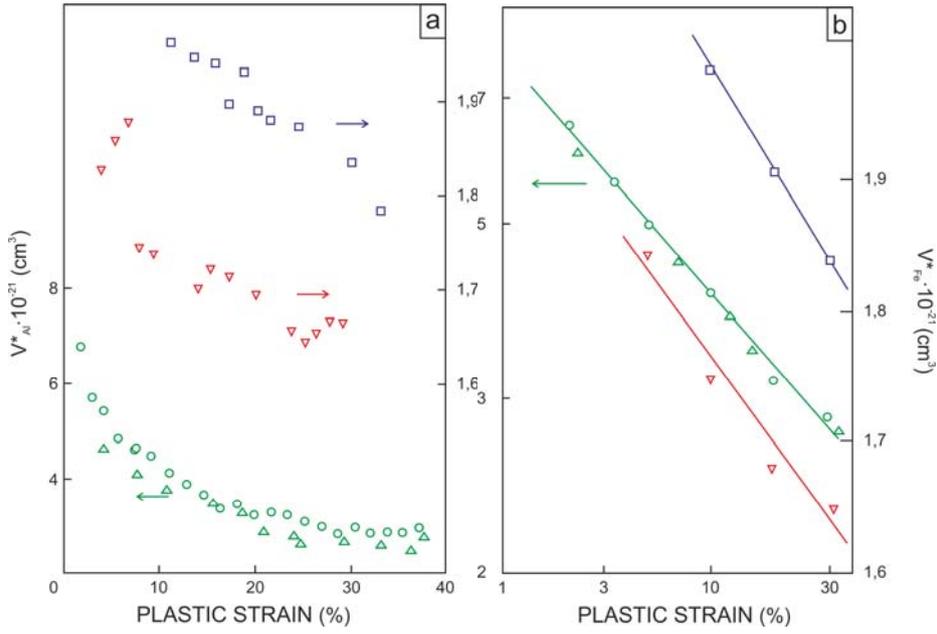

Figure 5. Activation volume, $V^*$, as a function of plastic strain ε for Fe, $T_{ann}= 900°$ C, (cubes); Fe, $T_{ann}=1150°$ C (inverse triangles) and Al [4] (triangles and circles) in linear (a) and logarithmic (b) scales. Triangles corresponds to Al without irradiation; circles –irradiation at φ = $5×10^{13}$ cm$^{-2}$.s$^{-1}$. The lines in (b) are to guide an eye.

Comparing dependencies of $δσ_φ$ and $V^*$ on ε (Figs. 3 and 5) it is possible to express $δσ_φ$ directly via $V^*$, as it is shown in Fig. 6 for Fe specimens ($T_{ann}=900°$ C) under different electron fluxes in comparison with Al specimens of different thickness. The most remarkable common peculiarity is that $δσ_φ$ *increases linearly* with decreasing activation volume $V^*$, which indicates unambiguously that irradiation makes the obstacles effectively smaller in size thus enhancing the dislocation unpinning and promoting slip. This important observation will be recalled in section 5 to verify the proposed RIS model.

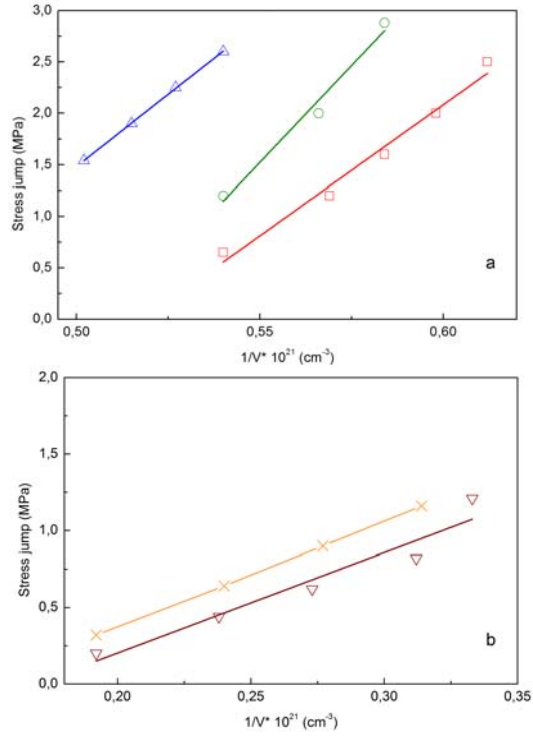

Figure 6. (a) Stress jump $\delta\sigma_\varphi$ as a function of the activation volume $V^*$ under electron beam with E = 0.8 MeV, $\varphi$ = 2.4×10$^{13}$ cm$^{-2}$.s$^{-1}$ (☐); 3.6×10$^{13}$ cm$^{-2}$.s$^{-1}$ (◯) and for Fe specimens (T$_{ann}$= 1150$^0$ C, $h$ =0.53 mm) - △; (b) Al specimens of thickness $h$ = 1.28 mm (▽) and $h$ = 0.58 mm (✕) under electron beam with E = 0.5 MeV, $\varphi$ = 5×10$^{13}$ cm$^{-2}$.s$^{-1}$. The lines are to guide the eye.

In the following section, we consider MD results on DB properties and their interaction with edge and screw dislocations in bcc Fe, which provide the basis for the construction of the RIS model.

## 3  DB properties and their interaction with edge and screw dislocations in bcc Fe

### 3.1 Existence of DBs in transition metals

DBs do not radiate their energy in the form of small-amplitude waves because they vibrate at frequencies outside the phonon spectrum of crystal [8-10]. DB frequency can leave the phonon spectrum when its amplitude is sufficiently large because the frequency of a nonlinear oscillator is amplitude-dependent. There are two types of nonlinearity, the hard-type and the soft-type. In the former (latter) case the DB frequency increases (decreases) with increase in its amplitude. In the case of the hard-type nonlinearity the DB frequency can be above the phonon spectrum. For the soft-type nonlinearity DBs can exist only if the phonon spectrum possesses a gap, which is the case e.g. in the crystals with the NaCl structure [10], diatomic crystals with large mass difference of atoms [11] and graphane (fully hydrogenated graphene) [12].

For a long time, it has been assumed that the softening of atomic bonds with increasing vibrational amplitude is a general property of crystals, which means that the oscillation frequency decreases with increasing amplitude. Therefore DBs with frequencies above the upper phonon frequency were not expected to exist. However, in 2011, Haas et al [14] have provided a new insight into this problem by demonstrating that the anharmonicity of metals appears to be very different from that of insulators. Since essential contribution to the *screening* of the atomic interactions in metals comes from free electrons at the Fermi surface, the ion-ion attractive force may acquire a nonmonotonic dependence on the atomic distance and may be enhanced resulting in an amplification of even anharmonicities for the resulting two-body potentials [14]. This effect can counteract the underlying softening associated with the bare potentials with a moderate increase of vibrational amplitudes to permit the existence of DBs with frequencies outside the phonon spectrum. MD simulations of lattice excitation in fcc nickel as well as in bcc niobium and iron using realistic many-body interatomic potentials have proven that stable high-frequency DBs do exist in these metals [14, 15]. Notably, the excitation energy of DBs can be relatively small (fractions of eV) as compared to the formation energy of a stable Frenkel pair in those metals (several eV). Moreover, it has been shown that DBs in Fe (and most likely in other transition metals) are highly mobile, hence can efficiently transfer a concentrated vibrational energy over large distances along close-packed crystallographic directions [15, 22].

Recently, a theoretical background has been proposed to ascribe the interaction of moving DBs with defects in metals to explain the anomalously accelerated chemical reactions in metals subjected to irradiation. Moving DBs are also termed 'quodons' – quasi-particles propagating along close-packed crystallographic directions[1]). Irradiation may cause continuous generation of quodons inside materials due to *external lattice excitation*, thus 'pumping' a material with a quodon gas [18, 19].

Evidence, provided by the atomistic simulations proving the existence of standing and moving DBs in metals, raises an important question about the interaction of these DBs with primary lattice defects and subsequent change of material's properties as a consequence. With the aid of modern *ab initio*-derived many body interatomic potentials (IAP), including 'magnetic' potential for bcc Fe, we investigate the interaction of DBs with an edge and screw dislocation in bcc Fe in the following sections.

---

[1] It should be noted that Russell and Eilbeck [23] have presented experimental evidence for the existence of quodons that propagate great distances in atomic-chain directions in crystals of muscovite, an insulating solid with a layered crystal structure. Specifically, when a crystal of muscovite was bombarded with alpha-particles at a given point at 300 K, atoms were ejected from remote points on another face of the crystal, lying in atomic chain directions at more than $10^7$ unit cells distance from the site of bombardment.

### *3.2 MD setup*

The behavior of standing and moving DBs has been studied using classical MD simulations in 3D periodic bcc Fe. To ensure that conclusions are independent of the choice of the particular cohesive model, we exploit two well spread interatomic potentials (IAPs) including that developed by Chamati et al. [24] (used earlier in ref. [15] to demonstrate the existence of a moving DB in Fe) and the well-known 'magnetic' potential for bcc Fe, developed by Dudarev and Derlet [25]. Although being semi-empirical, such IAPs are derived to account for the electronic charge distribution depending on the local atomic arrangement and are known to provide a good compromise between computationally expensive *ab initio* calculations and over-simplified pairwise potentials. Both IAPs have been widely and successfully used to model bulk and surface properties of bcc Fe as well as to study point-, extended- and interface-like lattice defects.

MD simulations were done in the virtual 3D-periodic crystals with three principal axes x, y and z oriented along the <111>, <-12-1> and <-101> directions, respectively. A DB was created in the crystal by providing initial displacement along x direction to six neighboring atoms selected in the centre following the procedure proposed by Hizhnyakov et al. [14, 15]. The key feature of the procedure is the initial displacement of the two adjacent atoms from their equilibrium position along the close <111> direction, which should oscillate in the *anti-phase mode* with respect to each other thus forming a stable DB, as shown in Fig. 7 (a). The initial offset displacement $d_0$ determines the DB amplitude and oscillation frequency and, ultimately, its lifetime. DBs can be excited in a narrow frequency band $(1 \div 1.4) \times 10^{13}$ s$^{-1}$ just above the Debye frequency of bcc Fe, and DB frequency grows with increasing amplitude as expected from the "hard" type anharmonicity of the considered vibrational mode. Application of a displacement larger than 0.45 Å generates a chain of *focusons*, while a displacement smaller than 0.27 Å does not provide enough potential energy for the two oscillators to initiate a stable DB and the atomic oscillations decay quickly by losing its energy to *phonons*. The most stable DBs can survive up to 400 oscillations, as shown in Fig. 7 (b), and ultimately decay in a stepwise quantum nature by generating bursts of phonons, as has been predicted by Hizhnyakov as early as in 1996 [26].

In order to quantify movement of the DB and its interaction with the lattice defects, the atomic positions along close-packed direction (111), and the deviations of potential energy of atoms from the initial values were analyzed.

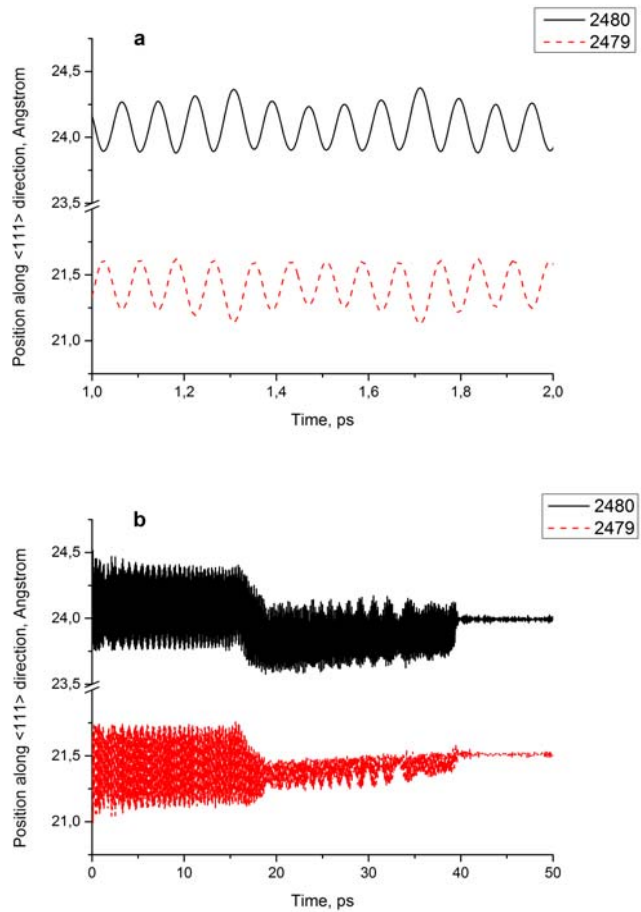

Figure 7. Oscillation of x coordinate of two neighboring atoms, 2480 and 2479 in a [111] row in Fe in a "standing" DB excited with $d_0 = 0.325$ Å using IAP from ref [24]. (a) Initial stage of DB evolution; (b) total lifespan of DB showing a stepwise quantum nature of its decay.

### 3.3 Interaction of moving DBs with dislocations

The movement of a DB can be induced by translational kinetic energy $E_{tr}$ given to the two central DB atoms in the same direction along the x-axis. Their velocities range from about 300 to 2000 m/s while travel distances range from several dozens to several hundreds of atomic spaces, depending on the $d_0$ and $E_{tr}$ [22]. Fig.2 (a) shows a DB approaching the atoms with index 3415 and 3416. The two atoms pulsate in the *anti-phase mode* for about 1ps (~10 oscillations) and then oscillations cease but they are resumed at the subsequent atoms along the x-axis. In this way, the DB moves at a speed of 2.14 km/s, i.e. about the half speed of sound in bcc Fe. The translational kinetic energy of the DB is about 0.54 eV, which is shared among two core atoms, giving 0.27 eV per atom. This is very close to the initial kinetic energy $E_{tr}$=0.3 eV transmitted to the atoms to initiate the DB movement. The deviation of the potential energy of the atoms from the ground state during the passage of the DB is presented in Fig.2b. The amplitude of the energy deviation can reach almost 1 eV. In an oversimplified 'thermodynamic' analogy, a moving DB can be viewed as an atom-size spot heated above 1000 K propagating through the crystal at sub-sonic speed. (see also movie added as supplementary material).

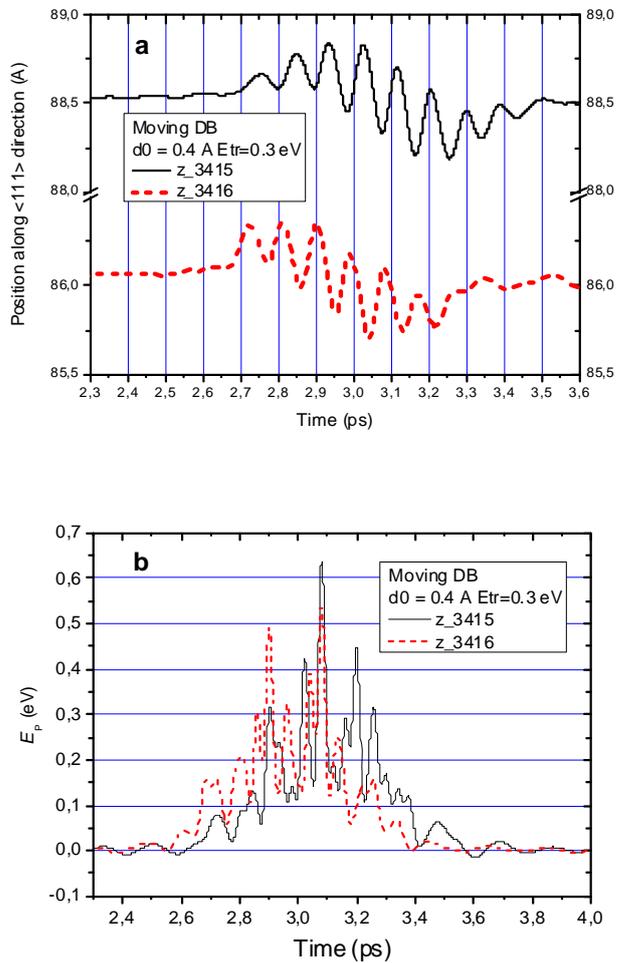

Figure 8 (a) Oscillation of x coordinate of two neighboring atoms, 3415 and 3416 in a [111] row in Fe during the passage of a moving DB ($d_0 = 0.4$ Å, $E_{tr}= 0.3$ eV); (b) deviation of the potential energy of the atoms from the ground state during the passage of DB. IAP is from ref [25].

The scattering of DBs on the core of an edge dislocation was studied for several aiming parameters. The DB passed through the dislocation in all the cases, except for the direct hit in the lower part of the dislocation core, as shown in the inset schematics in Fig.9. The only intensive DB scattering was registered for the interaction along the 'line 9'. The excitation of the three atoms forming the dislocation core is presented in Fig.9. Apparently, only the frontally hit atom 'B' exhibits essential vibration. The integral under the curve for the atom B is 0.15 eV, which is comparable with that for the near-vacancy atom A. The dislocation excitation time is about 2 ps, which is close to the excitation time for DB-vacancy or DB-free surface interaction evaluated in [22].

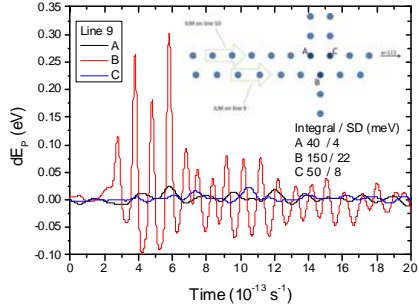

Figure 9. Variation of the potential energy of atoms surrounding a core of ½<111>{110} edge dislocation caused by the scattering of the DB ($d_0 = 0.4$ Å, $E_{tr}$=0.3 eV) moving along the 'Line 9'.

The scattering of the DBs on the core of a screw dislocation was also studied for DBs with $E_{tr}$ varied from 0.3 up to 0.6 eV. As in the case of the edge dislocation, we used a parallelepiped-like simulation box with the principal axes (x, y and z) oriented along [11-2], [1-1 0] and [111] directions, respectively. The dislocation line is placed at the center of the simulation box and periodic boundary conditions are applied along the dislocation line (i.e. along the [111] direction), atoms in several upper and lower layers normal to the [1-10] direction were fixed, while free surfaces are present normal to the [11-2] directions. The dislocation is created by applying the isotropic elastic solution of the displacement field to all atoms and then relaxing the atomic positions. The DB was created on the [-1-11] row, which belongs to the (1-10) plane and it is inclined to dislocation line at ~70° angle, as is shown on the MD snapshot in Fig.10(a). Initially, the DB is generated at a distance of about 5 nm from the dislocation line.

The evolution of the atomic positions in two (1-10) planes, cutting through the dislocation line, was followed by means of visualization, while the scattering intensity and period was measured by extracting the information on the atoms forming the dislocation core (see Fig.10(a)). The scattering of the DBs directly hitting the dislocation line resulted in a similar profile of $dE_P$ irrespective of the initial DB energy, see examples for $E_{tr}$= 0.35 and 0.55 eV in Fig.10(b). The excitation time amounted to ~0.4 ps (shorted than that for the edge dislocation) but the amplitude of $dE_P$ exceeds 1 eV, pointing to a different relaxation mechanism. Eventually, the scattering on the screw dislocation resulted in the creation of a kink and its propagation along the line, as revealed by visuzalization analysis (see movie added as supplementary material). As the studied crystal had periodic boundary condition along the dislocation line, the overall result was a displacement of the whole dislocation line by an elementary step. Performing the same simulations in the crystal with free surface along the dislocation line (i.e. [111] direction), we found that the same kink was created and moved to the surface. Hence, the scattering of the mobile DBs directly on the screw dislocation core results in the disturbance of the core which relaxes by the emission of the kink.

The interaction of the DBs that do not directly intersect the dislocation core produced negligible scattering basically showing that the capture radius of the screw dislocation is comparable to its atomic core size.

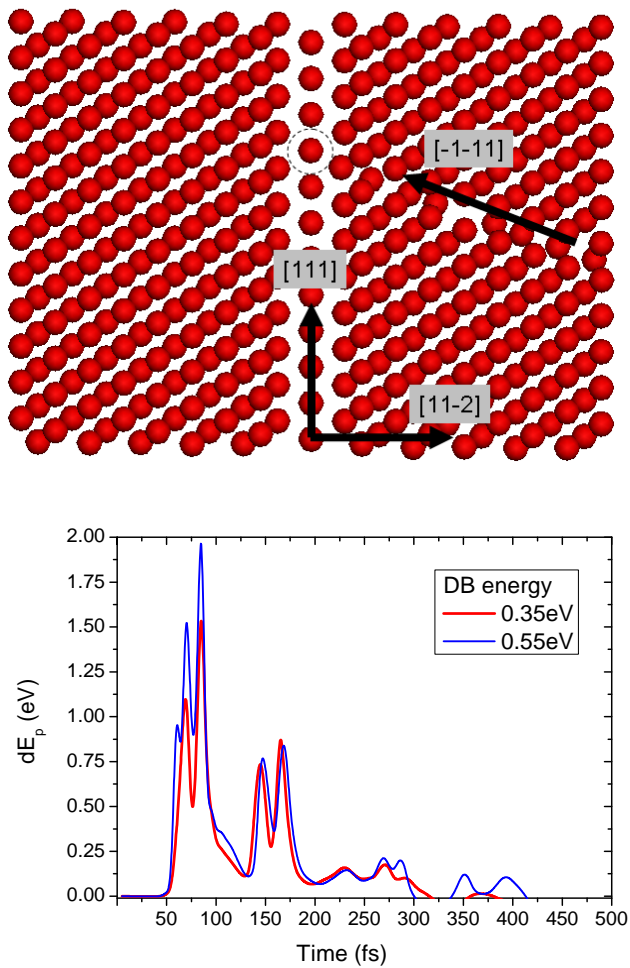

Figure 10. (a) (1-10) cross-section of the crystal containing the screw dislocation and DB moving towards its core. The dislocation core, aligned along [111] direction, is in the center of the figure. The evolution of the potential energy of an encircled atom is presented in fig. (b). The interaction of the DB with the dislocation core can be viewed on animation provided as on-line supplementary material.

DB excitations and scattering on the defects presented here as well as in refs. [14, 15, 22] were done in conditions imitating a periodic perfect crystal at zero temperature, i.e. when all other atoms were initially at their lattice positions and had zero initial velocities. This poses an important question about possible effect of finite lattice temperature and the crystal size on the robustness of the obtained DBs. This questions were investigated by Zhang and Douglas [27] as discussed in section 6.

## 4    Rate theory of DB excitation under thermal equilibrium and external driving

The rate equation for the concentration of DBs with energy $E$, $C_{DB}(E,t)$ can be written as follows [17]

$$\frac{\partial C_{DB}(E,t)}{\partial t} = K_{DB}(E) - \frac{C_{DB}(E,t)}{\tau_{DB}(E)},\tag{2}$$

where $K_B(E)$ is the rate of DBs generation with energy $E > E_{min}$ and $\tau_{DB}(E)$ is the DB lifetime. It has an obvious steady-state solution ($\partial C_{DB}(E,t)/\partial t = 0$):

$$C_{DB}(E) = K_{DB}(E)\tau_{DB}(E),\tag{3}$$

In the following sections we will consider the breather formation by thermal activation and then extend the model to non-equilibrium systems where lattice excitation provoked by irradiation generate additional DBs.

### 4.1    Thermal activation and DB lifetime

Following the direct observation of the generation of DBs in MD simulations at high temperature [11], we propose that its nucleation rate is given by Arrhenius law [7] as:

$$K_{DB}(E,T) = \omega_{DB}\exp\left(-\frac{E}{k_BT}\right),\tag{4}$$

where $\omega_{DB}$ is the attempt frequency that should be close to the DB frequency, and $E$ is the energy needed for the DB excitation, which may be called the DB energy. The breather lifetime has been proposed to be determined by a phenomenological law based on two basic principles [7]: (i) DBs have a minimum energy $E_{min}$, which means that there is always a non-zero energy gap for exciting a DB in two-dimensional and three-dimensional perfect lattices [7]. (ii) The lifetime of a breather grows with its energy as $\tau_{DB} = \tau_{DB}^0\left(\frac{E}{E_{min}} - 1\right)^z$, with $z$ and $\tau_B^0$ being constants, whence it follows that under thermal equilibrium, the DB energy distribution function $C_{DB}(E,T)$ and the mean number of breathers per site $n_{DB}(T)$ are given by:

$$C_{DB}(E,T) = \omega_{DB}\tau_{DB}\exp\left(-\frac{E}{k_BT}\right),\qquad(5)$$

$$n_{DB}(T) = \int_{E_{\min}}^{E_{\max}} C_{DB}(E,T)\,dE = \omega_{DB}\tau_{DB}^0 \frac{\exp\left(-\dfrac{E_{\min}}{k_BT}\right)}{\left(E_{\min}/k_BT\right)^{z+1}}\int_0^{\frac{E_{\max}-E_{\min}}{k_BT}} y^z \exp(-y)\,dy,\qquad(6)$$

Noting that $\Gamma(z+1,x) = \int_0^x y^z \exp(-y)\,dy$ is the second incomplete gamma function, eq. (5) can be written as [17]

$$n_{DB} = \omega_{DB}\tau_{DB}^0 \frac{\exp(-E_{\min}/k_BT)}{\left(E_{\min}/k_BT\right)^{z+1}}\Gamma\left(z+1,\frac{E_{\max}-E_{\min}}{k_BT}\right),\qquad(7)$$

It can be seen that the mean DB energy is higher than the averaged energy density (or temperature):

$$\langle E_{DB}\rangle = \frac{\displaystyle\int_{E_{\min}}^{E_{\max}} C_{DB}(E,T)E\,dE}{\displaystyle\int_{E_{\min}}^{E_{\max}} C_{DB}(E,T)\,dE}\xrightarrow{\;E_{\max}\gg E_{\min}\;}\left(\frac{E_{\min}}{k_BT}+z+1\right)\times k_bT\;,\qquad(8)$$

Following [11], one can assume that $E_{\min}/k_BT \approx 3$ and $\langle E_B\rangle \approx 5k_BT$, thus an estimate for $z \approx 1$, which corresponds to linear increase of the DB lifetime with energy.

### 4.2 External driving

Fluctuation activated nature of DB generation can be described in the framework of classical Kramers model, which is archetypal for investigations in reaction-rate theory [28]. The model considers a Brownian particle moving in a symmetric double-well potential $U(x)$, such as shown in Fig. 11 (a). The particle is subject to fluctuational forces that are, for example, induced by coupling to a heat bath. The fluctuational forces cause transitions between the neighboring potential wells with a rate given by the famous Kramers rate:

$$\dot{R}_K(E_0,T) = \omega_0 \exp(-E_0/k_BT),\qquad(9)$$

where $\omega_0$ is the attempt frequency and $E_0$ is the height of the potential barrier separating the two stable states, which, in the case of the fluctuational DB creation, corresponds to the minimum energy to be transferred to particular atoms to initiate a stable DB. Thus, the DB generation rate (3) is given by the Kramers rate: $K_{DB}(E,T) = \dot{R}_K(E,T)$.

In the presence of *periodic modulation* (driving) of the well depth (or the reaction barrier height) such as $U(x,t) = U(x) - V(x/x_m)\cos(\Omega t)$, the reaction rate $\dot{R}_K$ averaged over times exceeding the modulation period has been shown to increase according to the following equation [17]:

$$\left\langle \dot{R} \right\rangle_m = \dot{R}_K \, I_0\!\left( \frac{V}{k_b T} \right), \qquad (10)$$

where the amplification factor $I_0(x)$ is the zero order modified Bessel function of the first kind. Note that the amplification factor is determined by the ratio of the modulation amplitude $V$ to temperature, and it does not depend on the modulation frequency or the mean barrier height. Thus, although the periodic forcing may be too weak to induce *athermal* reaction (if $V < E_0$), it can amplify the average reaction rate drastically if the ratio $V/k_b T$ is high enough, as it is demonstrated in Fig. 11 (b).

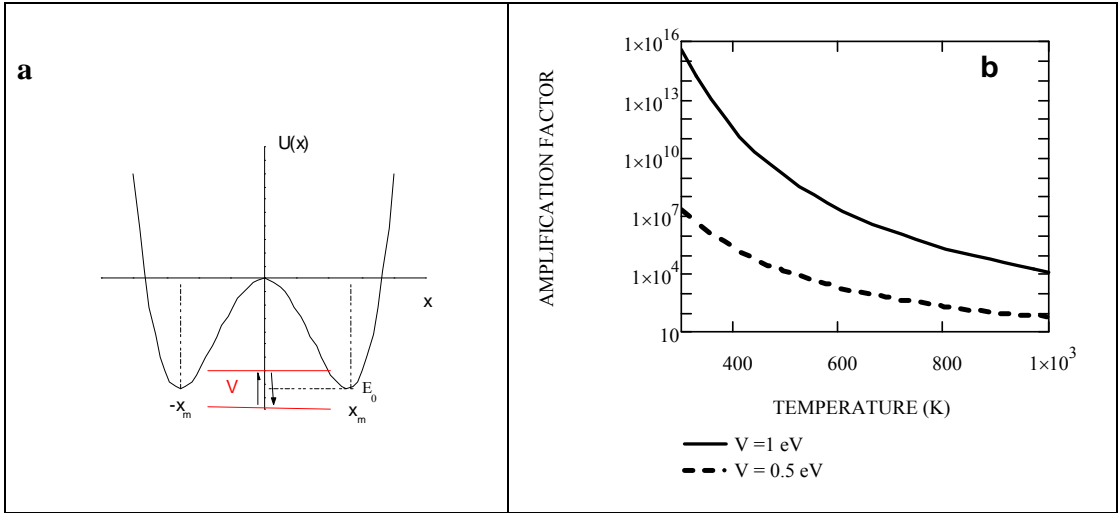

Figure 11. (a) Sketch of the double-well underline{potential landscape} with minima located at $\pm x_m$. These are stable states before and after reaction, separated by a potential "barrier" with the height changing periodically or stochastically within the V band. (b) Amplification factor, $I_0(V/k_B T)$, for the average escape rate of a thermalized Brownian particle from a periodically modulated potential barrier at different temperatures and modulation amplitudes, V [17].

Another mechanism of enhancing the DB creation rate is based on small *stochastic modulations* of the DB activation barriers caused by external driving. Stochastic driving has been shown to enhance the reaction rates via effective reduction of the underlying reaction barriers [18, 19] as:

$$\left\langle \dot{R} \right\rangle = \omega_0 \exp\!\left( -E_a^{DB} / k_b T \right), \; E_a^{DB} = E_0 - \frac{\left\langle V \right\rangle_{SD}^2}{2 k_b T}, \qquad (11)$$

where $\langle V \rangle_{SD}$ is the standard deviation of the potential energy of atoms surrounding the activation site.

In the present view, the DB creation is seen as a *chemical reaction* activated by thermally or externally induced fluctuations. In the following section we consider the reaction of unpinning of dislocations from obstacles, such as the crystal defects, within a similar framework.

## 5  Amplification of dislocation unpinning by moving DBs

As has been noted in section 2, an instant stress jump $\delta\sigma_\varphi$ at the turn on/off the beam is a purely *athermal effect* since it is occurs too fast to be explained by temperature excursion (due to beam). The effect enhances with increasing deformation level and electron flux as well as with decreasing specimen thickness, as shown in Figs. 3, 4, 6. An important observation is that $\delta\sigma_\varphi$ is inversely proportional to the activation volume $V^*$ (Fig. 6), which indicates unambiguously that irradiation *activates* the dislocation unpinning from the obstacles.

In order to evaluate the yield stress under a constant strain rate $\dot{\varepsilon}_{ex}$ one has to solve eq. (1) with respect to $\sigma$, which results in the well-known expression describing the yield stress dependence on temperature and the strain rate:

$$\sigma_T = \frac{E_a}{V_a} - \frac{k_b T}{V_a} \ln\left(\frac{\dot{\varepsilon}_0}{\dot{\varepsilon}_{ex}}\right), \tag{12}$$

To take into account effects caused by radiation-induced *moving DBs*, we remind that their scattering on dislocations leads to the atomic excitations at the dislocation cores (see Figs. 9, 10). This process should result in quasi-periodic modulations of the activation barriers for the dislocation movement. The amplitude of the quasi-periodic energy deviations $V_{ex}$ is within in the range of 0.1-1 eV with the excitation time $\tau_{ex}$ depending on the type of the obstacle and DB kinetic energy but typically not exceeding several picoseconds. In the modified Kramers model (see eq. 10), this energy deviation corresponds to the modulation of the unpinning activation barrier. Then, a *macroscopic* strain rate (averaged over a macroscopic number of obstacles and activated dislocation segments) will be determined by the following equation:

$$\langle \dot{\varepsilon}_{DB} \rangle_{macro} = \dot{\varepsilon}_T \left( 1 + \left\langle I_0 \left( \frac{V_{ex}}{k_b T} \right) \right\rangle \omega_{ex} \tau_{ex} \right), \tag{13}$$

where the brackets designate averaging over the excitation energies, and $\omega_{ex}$ is the mean number of excitations per obstacle per second caused by the flux of quodons, which is proportional to the quodon flux $\Phi_q$ and the cross-section of quodon-obstacle interaction $S_{ex}$ and is given by:

$$\omega_{ex} = \Phi_q S_{ex} \quad , \quad S_{ex} = b l_{ex}, \tag{14}$$

where, $l_{ex}$ is the characteristic distance along the dislocation line, from which the excitation caused by quodon scattering affects the obstacle. The flux of quodons produced by irradiation is proportional to the electron flux $\varphi$ and to the ratio of the electron energy $E_e$ to the mean quodon formation energy $E_q$:

$$\Phi_q = \varphi \frac{E_e}{E_q}, \tag{15}$$

whence the yield stress, *reduced under the DB-driven dislocation activation*, can be derived as follows:

$$\sigma_q = \frac{E_a}{V^*} - \frac{k_b T}{V^*} \ln\left(\frac{\dot{\varepsilon}_0}{\dot{\varepsilon}_{ex}}\left(1 + I_0\left(\frac{E_{ex}}{k_b T}\right)\omega_{ex}\tau_{ex}\right)\right), \tag{16}$$

which, in contrast to the thermally activated yield stress (see eq. 12) depends on the irradiation flux.

For polycrystalline materials, the above evaluated yield stress should be multiplied by the Taylor factor, $M$, ($M = 2.75$ for bcc metals; 3.10 for fcc metals).

The jump magnitude $|\delta\sigma_\varphi|$ is given by the difference between the expressions (12) and (16) for zero and non-zero irradiation flux $\varphi$. The instant yield stress jump at the turn on/off the beam occurs so quickly that the specimen temperature and activation volume remain constant. Accordingly, the classical expression (12), which does not depend on $\varphi$, would predict no RIS effect: $\left|\delta\sigma_\varphi\right|_T = 0$, while the expression (16) results in the following RIS effects for polycrystalline materials:

$$\left|\delta\sigma_\varphi\right|_q = M \frac{k_b T}{V^*} \ln\left(1 + I_0\left(\frac{E_{ex}}{k_b T}\right)\omega_{ex}\tau_{ex}\right), \tag{17}$$

One can see that the $\delta\sigma_\varphi$ value is inversely proportional to the activation volume, and it increases with increasing electron flux which agrees with experimental data, and allows one to evaluate the DB parameters from a quantitative comparison of the model with the data shown in Fig. 12. The red (online) dashed line corresponds to eq. (17) at irradiation parameters of Al specimen with a thickness of 0.58 mm (Fig. 6b), excitation energy of 0.38 eV and excitation lifetime of $10^{-11}$ s (Table 1), from which a constant of 0.7 MPa was subtracted to give the best fit to the experimental data ($\times$). The dotted line is obtained by subtracting from eq. (17) a constant equal to 0.9 MPa, which fits the experimental data for Al specimens with a larger thickness of 1.28 mm ($\triangledown$). It means that increasing specimen thickness beyond the electron penetration range (0.6 mm for E = 0.5 MeV) does not affect the quodon parameters (the angle of the line) but it decreases the RIS effect (the line shift).

The dash-dot lines (theory) and symbols (experiment) correspond to Fe specimens ($h$ = 0.53 mm) annealed at 900 °C under electron beam with E = 0.8 MeV and two electron fluxes φ = $2.4 \times 10^{13}$ cm$^{-2}$.s$^{-1}$ ($\square$); $3.6 \times 10^{13}$ cm$^{-2}$.s$^{-1}$($\bigcirc$). Fitting a stronger dependence on $V^*$ (higher line angle) requires one to assume higher excitation energy of 0.45 eV, while the positions of the lines have been set by subtracting from eq. (17) the constants equal to 13.5 and 12.3 MPa, respectively. For an initially less hard Fe specimens annealed at 1150 °C , the line position is set by subtracting from eq. (17) the constants 12 (solid blue line vs. $\triangle$). Thus, the angle of the lines is determined by the excitation energy and the electron flux, in agreement with the present model, while the line positions seem to be related to the specimen thickness and microstructure, which needs further investigation.

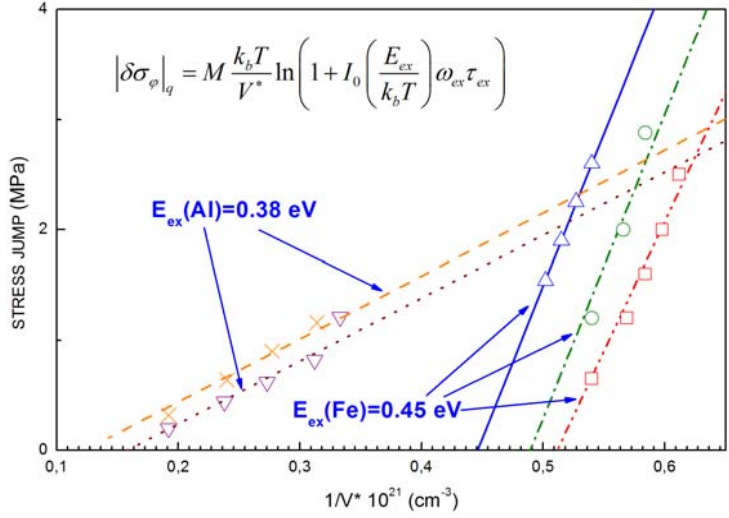

Figure 12. The yield stress jump, δσ_φ, as a function of the inverse activation volume, $V^*$, according to eq. (17) shown by the lines, compared to the experimental data for Al specimens $h$ = 1.28 mm ($\triangledown$) and $h$ = 0.58 mm ($\times$) at E = 0.5 MeV, φ = $5 \times 10^{13}$ cm$^{-2}$.s$^{-1}$ and for Fe specimens (T$_{ann}$=900° C) under electron beam with E = 0.8 MeV, φ = $2.4 \times 10^{13}$ cm$^{-2}$.s$^{-1}$ ($\square$); $3.6 \times 10^{13}$ cm$^{-2}$.s$^{-1}$ ($\bigcirc$); Fe specimens (T$_{ann}$=1150° C), $3.6 \times 10^{13}$ cm$^{-2}$.s$^{-1}$ ($\triangle$). Material and DB parameters used in calculations are presented in Table 1.

## 6    Discussion

The RIS effect demonstrated experimentally in the present paper for Fe and steel 20, as well as the previously studied RIS of Al and Cu, has been explained in the framework of a model based on the interaction of dislocations with moving DBs, a.k.a. quodons. Quodons can be generated by swift particles with energies above or below the threshold required for creation of stable displacements, i.e. Frenkel pairs, which explains why the RIS occurs under both sub- and over-threshold irradiation. In the latter case, the number of dpa produced during the time of experiment is very small (less than 10$^{-6}$ for the time to the specimen fracture) and does not affect the RIS significantly. It should be noted that direct scattering of swift electrons at dislocations also can facilitate their unpinning from obstacles. However, the flux of electrons is lower than the flux of quodons they can produce, by a factor of $\sim 10^5$, which practically rules out direct scattering of swift electrons at dislocations as a possible RIS mechanism.

Note that most of the experiments demonstrating RIS in Fe were conducted at temperatures above 300 K in polycrystalline metals. This poses an important question about possible effect of finite lattice temperature and the crystal size on the robustness of the obtained DBs. Recently, Zhang and Douglas [27] investigated interfacial dynamics of Ni nanoparticles at elevated temperatures exceeding 1000 K and discovered a string-like collective motion of surface atoms with energies in the eV range, i.e. exceeding the average lattice temperature by an order of magnitude. One of the most intriguing observations of this study was the propagation of the *breather excitations* along the strings, providing a possible mechanism for driving the correlated string-like atomic displacement movements. The authors conclude that these dynamic structures might be of crucial significance in *relation to catalysis*. High-temperature excitation of DBs has been demonstrated also for the two-dimensional crystal of $A_3B$ composition with long-range Morse interactions [11].

RIS effect is expected to operate also under neutron irradiation in reactor environment, since only a small part of the neutron energy is spent on stable displacements [19], while the major part of energy is dissipated into heat, or in other words, is spent on generation of phonons. As has been proposed in refs [18, 19], *quodons may be the transient form of the heat generation under irradiation*, which means that they are constantly generated by irradiation, and subsequently lose energy by generating phonons. This may change mechanical properties of materials under reactor conditions as compared to the surveillance specimens in *out-reactor tests* after equivalent irradiation dose, and so mechanical tests should be performed directly in a reactor to be meaningful. Such tests were scarce due to the high costs and technical problems, but they have been performed in a few cases described bellow.

Direct testing of mechanical properties under neutron irradiation in a fission reactor has been reported by Grynik and Karasev [29]. Polycrystalline Fe (99.99%) and its alloys Fe-$^{11}$B (0.001 wt%), Fe-$^{10}$B (0.003 wt%) and single crystal Fe-Si (3 wt%) have been mechanically tested in-situ under reactor irradiation at 10 MWt nuclear reactor at irradiation temperature $T_{irr} = (350 \pm 20)$ °C and the neutron flux of $10^{14}$ n/cm$^2$s. By means of internal friction measurements, the temperature dependence of the effective shear modulus, G, was measured under *in-situ irradiation* in the $T_{irr}$ range 350-700 °C, and compared to that measured *without irradiation* (prior and after irradiation to the same neutron fluence) in the T range 60-700 °C. It was found that G decreased by $\sim 10\%$ under in-situ irradiation and was back to its normal value after reactor was stopped, i.e. the effect was reversible.

The shear modulus is known to increase after considerable displacement damage due to the dislocation pinning at radiation-induced defects [29]. This *positive* change of the shear modulus is irreversible $\Delta G_{ir}>0$, in contrast to the reversible change of G observed under in-situ irradiation, which was negative, $\Delta G_{rev}<0$, and so it should be attributed to the *neutron-induced* acceleration of the dislocation unpinning from defects. This acceleration is a clear manifestation of the RIS effect under neutron irradiation. Similar RIS effect was observed in the earlier works by the Grynik and Karasev for Ni and Fe as noted in ref [29].

Another attempt to examine dynamic behavior of dislocations in the environment of fission neutrons was made by Singh et al [30] who measured experimentally the deformation behavior of pure iron and Fe-Cr alloy during uniaxial tensile tests performed directly in a fission reactor. Both in Fe-Cr alloy and pure iron, the slip systems in the planes with the Schmid factor value of almost zero were found to be activated during the in-reactor as well as the post-irradiation tensile tests. The author concluded that this had to be a specific effect of irradiation since normally in the unirradiated and deformed bcc crystals only two sets of dislocation walls were formed near the two highly stressed shear planes (i.e. with the top values of Schmid factor). However, it's rather difficult to obtain meaningful information concerning the RIS effect from this experiment. The point is that tensile tests were carried out at a constant strain rate of $10^{-7}$ s$^{-1}$ (three orders of magnitude lower than a typical rate of tensile tests). This rather low strain rate was chosen to ensure that the specimen should survive the in-reactor deformation for long enough time to accumulate a displacement dose level of about 0.1 dpa. As a result, two opposite mechanisms, namely, a typical radiation-induced hardening (RIH) and *in-situ* RIS were operating simultaneously, making the interpretation of the results obscure. Definitely, this area needs much more experimental and modeling efforts in order to forecast the service life of reactor structural materials with account of both RIH and RIS effects.

## 7    Conclusions

The present experiments were designed to allow comparison between sub-threshold and over-threshold electron irradiation of bcc and fcc metals, which did not show any significant difference.

The *radiation-induced softening* (RIS) effect was demonstrated for technically pure Fe as well as for previously studied Al and Cu. It is represented by (i) reversible decrease of the yield stress by ~ 1% at the moment of switching on electron beam and (ii) by irreversible decrease of the yield stress by ~ 10% under continuous irradiation up to the material fracture.

The RIS effect on the elongation to fracture of Fe appears to be opposite to that for Al and Cu, where electron irradiation increases elongation to fracture by ~25-30%. In contrast to that, the elongation to fracture of Fe specimens was *decreased* by irradiation by ~20%, which is significant reduction as compared to the result obtained by the standard test. This discrepancy may have important implications regarding the programmes forecasting the service lifetime of Fe-based structural steels.

On the theoretical part, a model of the RIS was developed based on a rate theory modified to account for the interaction of dislocations with *moving discrete breathers* produced by irradiating particles. In agreement with experimental data, the RIS increases with increasing *irradiation flux*, and it is inversely proportional to the *activation volume*.

We may conclude that the radiation-induced formation of DBs may change mechanical properties of materials under reactor conditions as compared to the surveillance specimens in out-reactor tests after equivalent irradiation dose. The RIS phenomenon needs further investigations due to its importance for the adequate qualification of the mechanical performance of the materials under reactor operating conditions.

With respect to other technological applications of DB creation in Fe-based alloys and steels, the above demonstrated DB-dislocation interaction, as well as the DB-defect interaction discussed in ref [22], should result in the modification of such important properties as ductility, creep resistance, swelling resistance, phase stability and others diffusion-limited processes responsible for the degradation of mechanical properties of structural steels under ageing and/or irradiation [16-19].

**Acknowledgements**

This study has been supported by the STCU Grants # 5497. DT acknowledge the support of the EUROfusion consortium within the Enabling Research grant and travel fellowship from Erasmus Mundus Fusion EP.

Table 1. Material and DB parameters used in calculations

| Parameter | Value |
| --- | --- |
| Atomic spacing, $b$ (nm) | 0.323 |
| Electron energy, $E_e$ (MeV) | 0.5÷0.8 |
| Electron flux, $\varphi$ (cm$^{-2}$s$^{-1}$) | (2.4÷5)×10$^{13}$ |
| Mean quodon energy, $E_q$ (eV) | 1 |
| Mean excitation energy, $E_{ex}$ (eV) | 0.38 (Al); 0.45 (Fe) |
| Mean excitation time, $\tau_{ex}$, (s) | 10$^{-11}$ |
| Mean excitation distance, $l_{ex}$ (b) | 10 |